\documentclass[journal=ancac3,manuscript=article]{achemso}

\usepackage[version=3]{mhchem} 

\usepackage{graphicx}
\usepackage[export]{adjustbox}
\usepackage{amsmath}
\usepackage{subcaption}
\usepackage{afterpage}
\usepackage{mathtools}
\usepackage{natbib}
\usepackage{hyperref}
\usepackage{xcolor}

\DeclarePairedDelimiter\abs{\lvert}{\rvert}

\author{Tobias Vogl}
\email{Tobias.Vogl@anu.edu.au}
\affiliation{Centre for Quantum Computation and Communication Technology, Department of Quantum Science, Research School of Physics and Engineering, The Australian National University, Acton ACT 2601, Australia}
\author{Marcus W. Doherty}
\affiliation{Laser Physics Centre, Research School of Physics and Engineering, The Australian National University, Acton, ACT 2601, Australia}
\author{Ben C. Buchler}
\affiliation{Centre for Quantum Computation and Communication Technology, Department of Quantum Science, Research School of Physics and Engineering, The Australian National University, Acton ACT 2601, Australia}
\author{Yuerui Lu}
\affiliation{Centre for Quantum Computation and Communication Technology, Research School of Electrical, Energy and Materials Engineering, The Australian National University, Acton ACT 2601, Australia}
\author{Ping Koy Lam}
\email{Ping.Lam@anu.edu.au}
\affiliation{Centre for Quantum Computation and Communication Technology, Department of Quantum Science, Research School of Physics and Engineering, The Australian National University, Acton ACT 2601, Australia}

\title{Atomic Localization of Quantum Emitters in Multilayer Hexagonal Boron Nitride}


\begin{document}
\begin{abstract}
The recent discovery of single-photon emitting defects hosted by the two-dimensional wide band gap semiconductor hexagonal boron nitride (hBN) has inspired a great number of experiments. Key characteristics of these quantum emitters are their capability to operate at room temperature with a high luminosity. In spite of large theoretical and experimental research efforts, the exact nature of the emission remains unresolved. In this work we utilize layer-by-layer etching of multilayer hBN to localize the quantum emitters with atomic precision. Our results suggest the position of the emitters correlates with the fabrication method: emitters formed under plasma treatment are always in close proximity to the crystal surface, while emitters created under electron irradiation are distributed randomly throughout the entire crystal. This disparity could be traced back to the lower kinetic energy of the ions in the plasma compared to the kinetic energy of the electrons in the particle accelerator. The emitter distance to the surface also correlates with the excited state lifetime: near-surface emitters have a shorter compared to emitters deep within the crystal. Finite-difference time-domain and density functional theory simulations show that optical and electronic effects are not responsible for this difference, indicating effects such as coupling to surface defects or phonons might cause the reduced lifetime. Our results pave a way toward identification of the defect, as well as engineering the emitter properties.
\end{abstract}

\clearpage
The recent discovery of quantum emitters in two-dimensional (2D) materials attracted considerable attention, due to their applications in photonic quantum technologies\cite{10.1038/nphoton.2009.229}. These include unconditionally secure communication\cite{10.1103/RevModPhys.74.145}, quantum simulators\cite{10.1038/nphys2253} and quantum computing\cite{10.1038/nature08812}, which fueled the development of single-photon sources (SPSs). In contrast to their counterparts in 3D, quantum emitters hosted by 2D lattices are not surrounded by any high refractive index medium. This eliminates total internal and Fresnel reflection of emitted single-photons, making it possible to have intrinsically near-ideal extraction efficiency. Quantum emission has been reported from a diversity of materials, in semiconducting transition metal dichalcogenides (TMDs)\cite{10.1364/OPTICA.2.000347,nnano.2015.60,nnano.2015.67,nnano.2015.75,nnano.2015.79,10.1038/ncomms12978,10.1063/1.4945268,arXiv:1901.01042} and insulating hexagonal boron nitride (hBN)\cite{nnano.2015.242}. The large band gap of the latter even allows to resolve the zero phonon line (ZPL) at room temperature and thwarts non-radiative recombination of the localized exciton. Thus, single-photon emitters in hBN have an intrinsically high quantum efficiency which leads to significantly brighter emission\cite{nnano.2015.242,doi:10.1021/acs.nanolett.7b00444}. In addition, single-photon sources based on hBN are suitable for many practical field applications due to their resistance to ionizing radiation\cite{10.1038/s41467-019-09219-5}, temperature stability over a huge range spanning 800$\,$K\cite{PhysRevB.98.081414,10.1021/acsphotonics.7b00086}, long-term operation\cite{10.1021/acsphotonics.8b00127} and capabilities for integration with photonic networks\cite{10.1021/acsphotonics.7b00025,10.1088/1361-6463/aa7839}, as well as easy handling. While these emitters can occur naturally\cite{nnano.2015.242}, it is common to enhance the defect formation synthetically through chemical\cite{10.1021/acs.nanolett.6b03268} or plasma etching\cite{10.1021/acsphotonics.8b00127,10.1039/C7NR08222C}, $\gamma$-ray\cite{10.1038/s41467-019-09219-5}, ion\cite{10.1021/acsami.6b09875} and electron irradiation\cite{10.1021/acsami.6b09875,10.1021/acsami.8b07506} or near-deterministic stress-induced activation\cite{10.1364/OPTICA.5.001128}.\\
\indent The generally accepted model for the single-photon emission is based on a localized exciton. These fluorescent point-like defects introduce trap states into the electronic band gap, acting thus as an effective two-level system. In defiance of several attempts to identify the origin of the fluorescence using group theory and \textit{ab inito} density functional theory (DFT) calculations\cite{10.1039/C7NR04270A,10.1021/acsphotonics.7b01442,PhysRevB.97.064101}, the exact nature of the defects remains controversial. Possible defect candidates include the C$_\text{B}$V$_\text{N}$, V$_\text{B}$C$_\text{N}$, V$_\text{N}$N$_\text{B}$ and V$_\text{B}$ defects. It was recently noted, however, that widely used generalized gradient functionals can perform poorly and lead to misassignment of the defect states, hence, hybrid or long-range corrected functionals should be applied\cite{doi:10.1021/acs.jctc.7b01072}. Moreover, DFT calculations often assume monolayered supercells due to the exponential scaling with the number of atoms and limited computational resources, while most experimental works involve multilayer hBN. For yet not fully understood reasons, the optical emission signatures of quantum emitters hosted by mono- and multilayer hBN differ substantially\cite{nnano.2015.242}.\\
\indent On the experimental side, research efforts toward the identification\cite{arXiv:1811.05924,arXiv:1901.05952} are hampered by the strongly varying optical emission properties. These vary not only from defect to defect on different hBN crystals, but also for defects on the same host crystal. ZPLs have been reported in the UV\cite{10.1021/acs.nanolett.6b01368} and in the the visible spectrum from 550 to 800$\,$nm\cite{PhysRevB.98.081414,10.1021/acsphotonics.8b00127,10.1021/acsnano.6b03602,10.1021/acsphotonics.6b00736} and the excited state lifetimes vary from 20$\,$ns down to 0.3$\,$ns\cite{10.1021/acsphotonics.7b00025,10.1021/acsphotonics.8b00127}. A conclusive explanation for this requires additional experimental analysis. What is definitely known is the power saturation behavior is that of an idealized two- or multi-level system and the emitters exhibit an in-plane dipole. This indicates a low symmetry in-plane defect that is potentially comprised of vacancies and impurities.\\
\indent The variations in ZPL position cannot be explained alone by local strain in the crystal environment. The shifts caused by strain are too small to account for the variety of ZPLs\cite{10.1038/s41467-017-00810-2}. Of particular note is that the ZPLs seem to bunch in groups around 560$\,$nm\cite{10.1021/acsphotonics.8b00127}, 580$\,$nm\cite{arXiv:1902.07932}, 640$\,$nm\cite{nnano.2015.242,10.1021/acsnano.6b03602} and 714$\,$nm\cite{10.1021/acsnano.6b03602}. We define these as groups 1 through 4, respectively. It is believed that a different point-like defect is responsible for each group with the crystal lattice locally strained or changed otherwise, thus explaining the spread around these wavelengths. Shifts of the transition line caused by different isotopes would be much smaller than the emission linewidth. The vibronic bandshape of most defects is very similar, indicating that they have the same symmetry group. We note that there are occasional ZPLs falling into neither of these categories. It is likely that these originate from surface contaminants. Moreover, the bandshape of these differ from the bandshape typical for other emitters in the three groups, which supports this conjecture.\\
\indent Using super-resolution techniques, these defects have been localized in 2D with sub-diffraction resolution\cite{10.1038/s41467-018-03290-0}. The direct imaging on the atomic scale using high-resolution scanning transmission electron microscopy (STEM) is limited to a few layers, as the images contain information from all layers (essentially being a projection of all layers onto 2D). One way around this is to use a more advanced method like high-angle annular dark-field imaging (HAADF), with which it is possible to detect the presence of a vacancy within a few layers (maybe up to 3-5 layers). A vacancy would change the detected intensity by changing the scattering probability locally, and thus this would reveal such a defect with the exact location in the XY plane. However, this still does not contain any information about the Z direction. Recently, a method to correlate optical and electron characterizations of quantum emitters in very thin hBN was demonstrated\cite{arXiv:1901.05952}. This method, however, also yields no information about the Z direction. In addition, detecting the presence of a vacancy using HAADF cannot be used on thicker crystals, because the intensity contrast would be too low.\\
\indent In this work, we localize the quantum emitters hosted by multilayer hBN in the third dimension with atomic precision. We develop deterministic layer-by-layer plasma etching of hBN. This way we can remove a single hBN monolayer at a time and check \textit{ex-situ} when the defect disappears. We thereby measure the precise distance of the emitter from the surface of the host crystal. While this is a destructive technique, it allows us to extract the exact number of layers in which the defect was located. Repeating our experiment for many defects allows us to generate sufficient statistics. We also model photophysical properties theoretically with finite-difference time-domain simulations and density functional theory.

\section*{Results and Discussion}
\subsection*{Layer-by-layer etching of hBN}
\begin{figure*}[t!]
\centering
  \includegraphics[width=0.58\linewidth,keepaspectratio,valign=t]{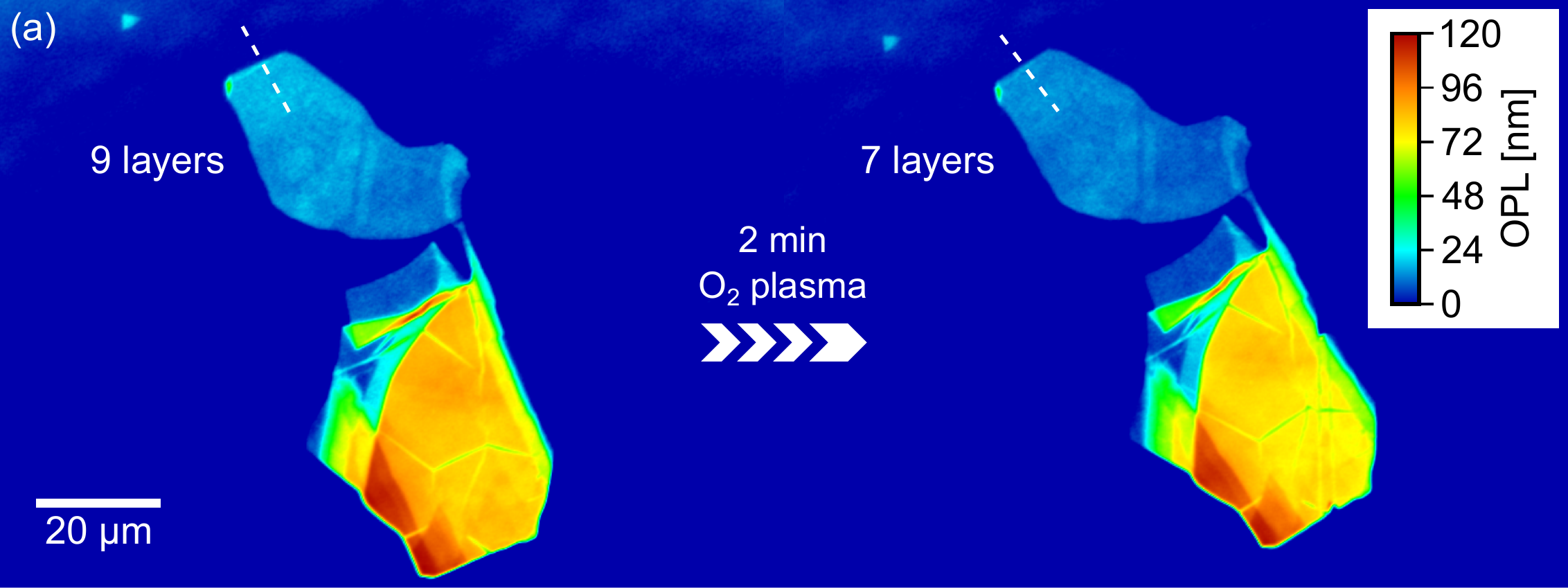}
  \includegraphics[width=0.327\linewidth,keepaspectratio,valign=t]{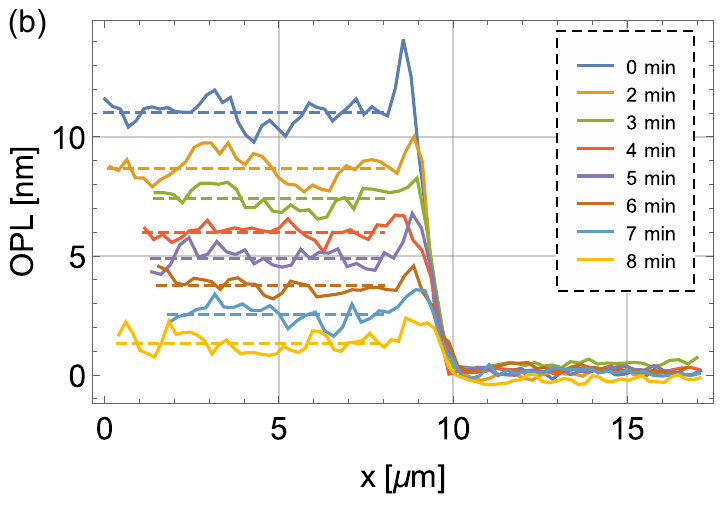}\\
  \includegraphics[width=0.327\linewidth,keepaspectratio,valign=t]{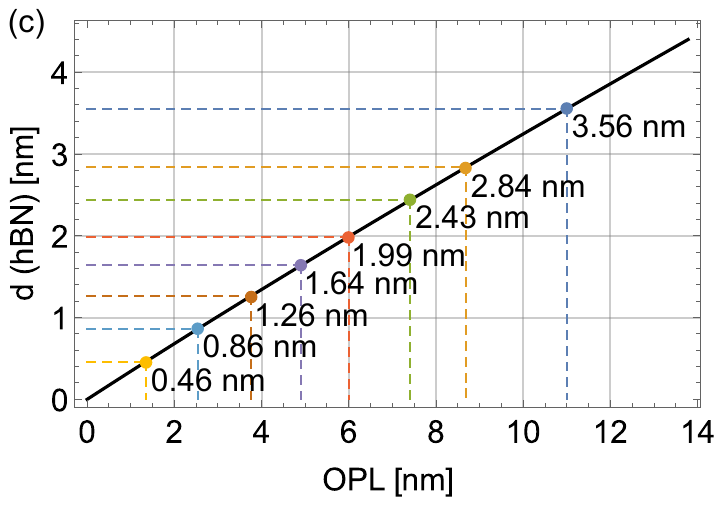}
  \includegraphics[width=0.327\linewidth,keepaspectratio,valign=t]{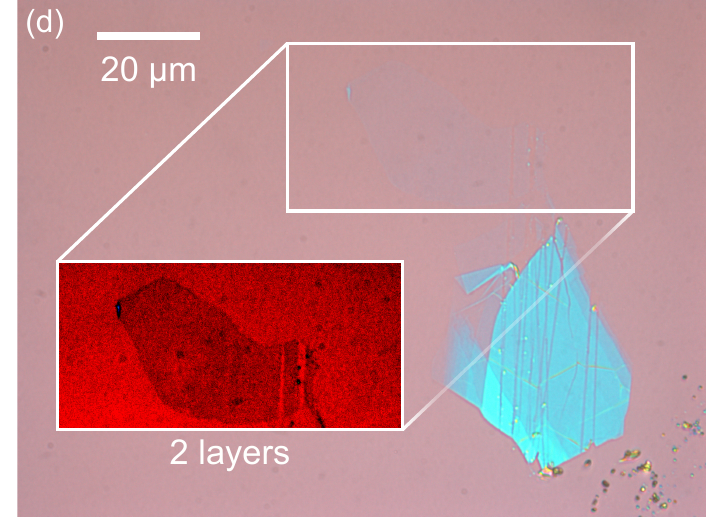}
  \includegraphics[width=0.327\linewidth,keepaspectratio,valign=t]{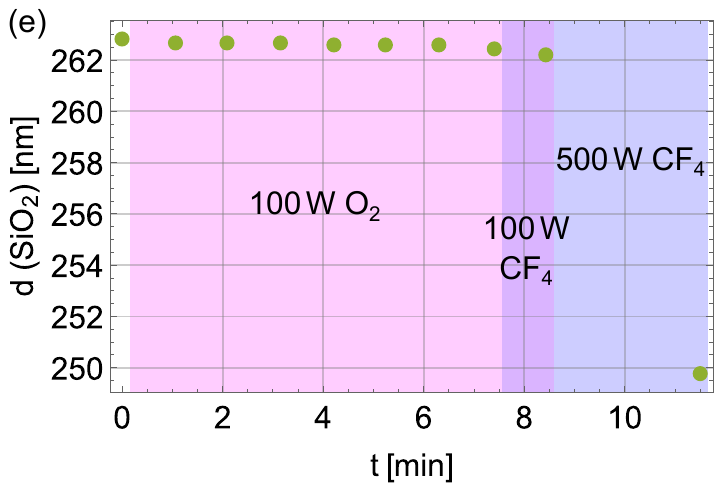}
\caption{Layer-by-layer etching of hBN. (a) PSI image of an hBN flake prior to any plasma treatment (left) and after 2$\,$min at 100$\,$W of oxygen plasma treatment (right). The thickness of the thin flake at the top is reduced from 9 to 7 atomic layers. The white dashed lines show the direction at which the traces in (b) are measured. (b) Optical path length difference along the white lines in (a) measured \textit{ex-situ} after each plasma etching step. The dashed lines denote the average. The start point of each is not equal. (c) RCWA simulation of the OPL difference for hBN on 262$\,$nm SiO$_2$ on Si (black line). The points visualize how the measured OPL can be converted into physical thickness of the flake. The physical thickness for each measured OPL is displayed in black next to the corresponding data point. (d) Microscope image ($1000\times$ magnification) of an hBN flake after 7$\,$min at 100$\,$W of oxygen plasma treatment. The crystal consists only of two atomic layers (for clarity the bilayer is shown). The inset shows a strongly contrast-enhanced image of the crystal. (e) Thickness of the SiO$_2$ layer on the Si substrate measured \textit{ex-situ} after each plasma etching step. After 7$\,$min at 100$\,$W (O$_2$), the thickness changed only marginally, by less than 0.22$\,$nm. After one additional minute at 100$\,$ (CF$_4$), the thickness further decreased by 0.22$\,$nm. After three additional minutes at 500$\,$W (CF$_4$) in the plasma field maximum, the thinning was substantial with 12.49$\,$nm decrease. The error bars are shorter than the size of the symbols. A significant change in the SiO$_2$ thickness would change the OPL.}
\label{fig:1}
\end{figure*}

Our approach to extract the location of the defects in the Z direction is to selectively remove one hBN monolayer at a time and check after each step, if the defect is still present. We first developed the layer-by-layer etching of hBN using an oxygen plasma. We note that similar etching of hBN on the atomic scale was reported recently using an argon plasma\cite{10.1039/C8NR02451K}. While this is an important milestone, however, Park \textit{et al.} etched $\sim 20$ layers at a time and scaled this down to monolayer etching\cite{10.1039/C8NR02451K}. Nevertheless, with this technique as well as our method (see below), it is possible to fabricate large hBN monolayers. These are very difficult to obtain using mechanical exfoliation alone, due to the poor optical contrast of hBN, which has a zero-crossing in the visible spectrum\cite{doi:10.1002/smll.201001628}.\\
\indent We mechanically exfoliated hexagonal boron nitride from bulk crystal onto a viscoelastic polymer. Thin, but still several nm thick hBN flakes were selected by optical contrast for dry transfer to a Si substrate terminated with a layer of thermally grown SiO$_2$ (262$\,$nm). For the etching we used an oxygen plasma generated from a microwave field and empirically optimized the plasma parameters (see Methods). The crystal thickness after each successive etching step is measured with a phase-shift interferometer (PSI), which is a much faster method than using an atomic force microscope (AFM) at the cost of a lower lateral resolution. Figure \ref{fig:1}(a) shows the PSI image prior to any plasma treatment and after 2$\,$ min of etching time, where the crystal thickness decreased. The top flake consists of 9 and 7 atomic layers, respectively. The optical path length (OPL) difference between the substrate and the crystal (measured along the white dashed lines in Figure \ref{fig:1}(a)) at a PSI wavelength of $\lambda=532\,$nm after each cumulative etching step is shown in Figure \ref{fig:1}(b). It can be seen that the etched thickness is linear with time. The OPL can be converted to physical thickness using rigorous coupled-wave analysis (RCWA) simulations\cite{10.1038/lsa.2016.46}, as shown in Figure \ref{fig:1}(c). The simulations assume the refractive index of hBN to be 1.849, which was extracted by fitting an RCWA model to data pairs consisting of AFM and PSI measurements. It is worth noting that the relation between OPL and physical thickness $d$ is nonlinear for large OPLs. The data points in Figure \ref{fig:1}(c) correspond to the PSI measurements (colored accordingly). Since the physical thickness of hBN is $0.4-0.45\,$nm per layer\cite{doi:10.1002/adfm.201504606}, we can extract that the crystal presented here was etched layer-by-layer from 9 layers to monolayer, with an etching rate of 1 layer per 63$\,$s. A microscope image with an artificially-enhanced optical contrast of the bilayer is shown in Figure \ref{fig:1}(d). At the optimized plasma conditions, this atomic layer-by-layer etching is highly reliable, with no fails (i.e.\ 0 or 2 layers etched) out of 31 runs. Moreover, we used the same technique on TMDs without failures and the method was also used for precise layer-by-layer thinning of black phosphorus\cite{10.1038/ncomms10450} or MoS$_2$\cite{10.1038/srep19945}. Assuming the failure probability to be $\leq 0.1\%$ would reproduce our etching success of hBN with a high probability of 96.9$\%$. Deviating from the ideal plasma conditions (63$\,$s etching time, for all details see Methods) results in process failures. This is evident by the fact that reducing the etching time by 10$\,$s resulted into 2 out of 6 crystals not being etched and increasing the etching time by 10$\,$s resulted in two layers being etched in 1 out of 3 cases (see Supporting Information, Figure S1). The reason why multiple layers can be etched without doubling the etching time is because it takes some time to start cracking the bonds, once that process starts, a faster etching rate can be achieved.\\
\indent It is important to note that the plasma may damage the substrate. The OPL is dependent on the SiO$_2$ thickness and the RCWA simulations assume this to be fixed. We checked the thickness of a SiO$_2$ layer \textit{ex-situ} after each etching step using variable angle spectroscopic ellipsometry (VASE). After 7$\,$min at 100$\,$W of cumulative plasma treatment, the thickness of a SiO$_2$ layer decreased from 262.68(1) to 262.46(1)$\,$nm (see Figure \ref{fig:1}(e)), so on average the SiO$_2$ thinning is $0.03\,$nm per step. According to the RCWA simulations such substrate thickness difference results in a change of the OPL much smaller than the resolution of the PSI (0.1$\,$nm). Therefore, we can neglect this effect. This is, however, in general dependent on the type of plasma. For a comparison: using a CF$_4$ plasma at 100$\,$W for 1$\,$min results in a thickness change of 0.22$\,$nm of the SiO$_2$ and using a CF$_4$ plasma at 500$\,$W for 3$\,$min in the plasma field maximum (see Methods) etches 12.49$\,$nm.

\subsection*{Creation of quantum emitters}
\begin{figure*}[t!]
\centering
  \includegraphics[width=0.327\linewidth,keepaspectratio,valign=t]{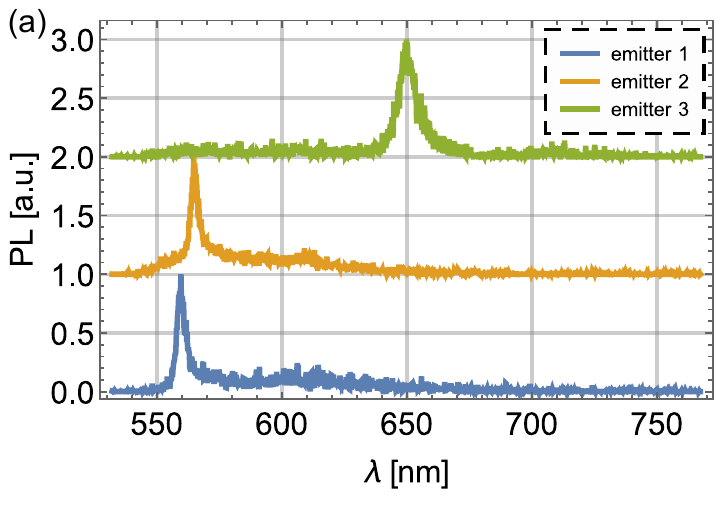}
  \includegraphics[width=0.327\linewidth,keepaspectratio,valign=t]{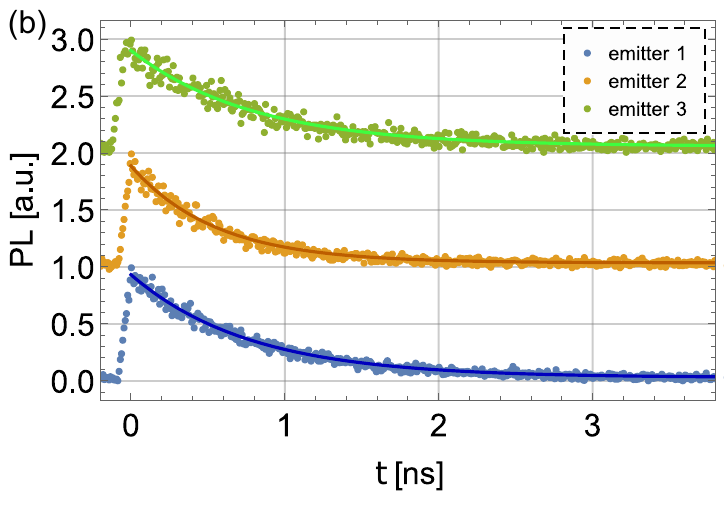}
  \includegraphics[width=0.327\linewidth,keepaspectratio,valign=t]{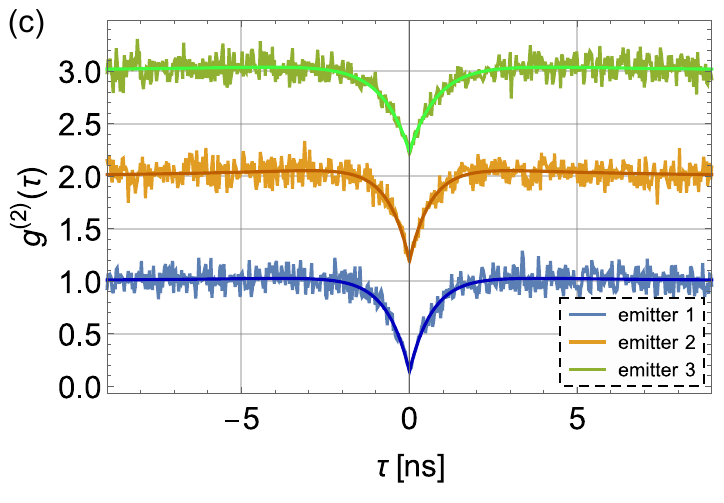}\\
  \includegraphics[width=0.327\linewidth,keepaspectratio,valign=t]{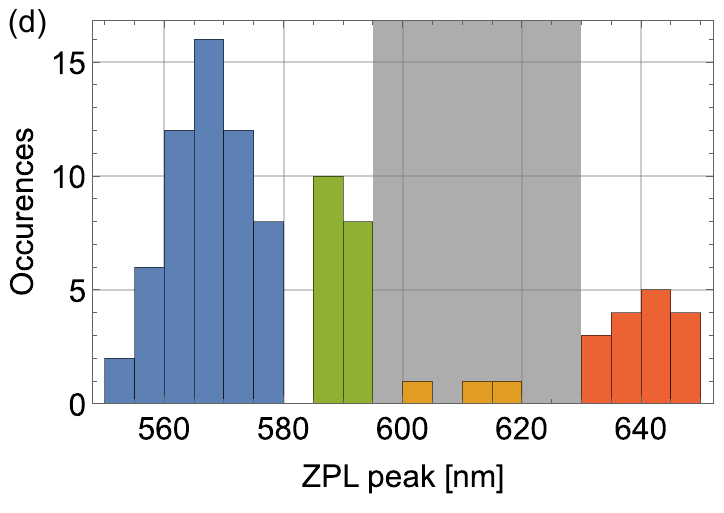}
  \includegraphics[width=0.327\linewidth,keepaspectratio,valign=t]{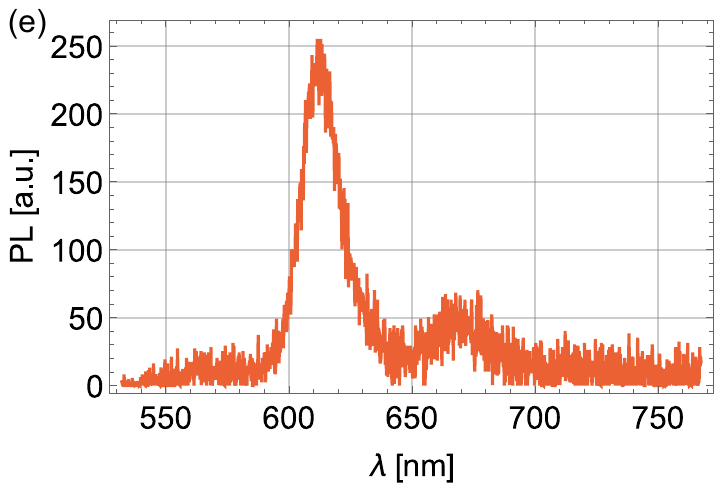}
\caption{Photophysics of the emitters. (a) Normalized spectra (vertically offset for clarity) of 3 sample emitters with their ZPLs at 559.78(7), 565.15(6) and 650.16(7)$\,$nm. Their corresponding Lorentzian linewidths are 2.24(10), 2.51(9) and 4.39(9)$\,$nm, respectively. (b) Time-resolved photoluminescence reveals a single-exponential decay of the excited state population with lifetimes 770(7), 549(7) and 794(13)$\,$ps for the emitters, respectively. The data is normalized and vertically offset for clarity. (c) The second-order correlation function dips to 0.142(37), 0.196(53) and 0.234(44) at zero time delay (obtained from fits). There was no background correction applied. The re-emission peaks are present, but not visible on the scales displayed. The data is normalized such that $g^{(2)}(\tau\rightarrow\infty)=1$ and vertically offset for clarity. (d) Histogram of the distribution of zero phonon lines from 93 defects. The ZPLs bunch around 560$\,$nm (group 1, blue), 590$\,$nm (group 2, green) and 640$\,$nm (group 3, red). It is believed that defects falling into neither of these categories (excluded area, grayed out) originates from surface contaminants. (e) Sample spectrum of such an emitter from the excluded area in (d). The emission of these emitters is typically comparably weak and broad.}
\label{fig:2}
\end{figure*}

The fabrication of multilayer hBN flakes for hosting single-photon emitter is similar to the procedure above. After transfer to the substrate, the flakes are treated with an oxygen plasma at different conditions and successively annealed in a rapid thermal annealer\cite{10.1021/acsphotonics.8b00127}. To locate the defects each flake is scanned in a custom-built confocal micro-photoluminescence ($\mu$PL) system with a resolution ranging from 0.2 or $1\,\mu$m. The pump laser, with its wavelength at 522$\,$nm, is blocked by a long-pass filter and the emission is collected in-reflection. The defects almost exclusively occur at the edges of the host crystal flakes, due to a lower defect formation energy at these locations. Defects can, however, also form along crystal cracks within the flake. The defect formation energy there is lower as well. The spectra of three sample emitters are shown in Figure \ref{fig:2}(a), which have their ZPLs at 559.78(7), 565.15(6) and 650.16(7)$\,$nm and Lorentzian linewidths of 2.24(10), 2.51(9) and 4.39(9)$\,$nm, respectively. All sample emitters presented here emit more than 80$\%$ of their photoluminescence (PL) into the ZPL, which allows for a high quantum efficiency. Time-resolved photoluminescence reveals a single-exponential decay of the excited state population for each defect with lifetimes 770(7), 549(7) and 794(13)$\,$ps, respectively (see Figure \ref{fig:2}(b)). The excitation laser is pulsed at a repetition rate of 20.8$\,$MHz and a pulse length of 300$\,$fs. While this allows for high peak intensities, two-photon absorption of the band gap of hBN is still impossible, because $E_\text{hBN}=6\,\text{eV}>2\times 2.38\,\text{eV}=2\times E_\text{laser}$. To prove that the localized exciton emits indeed non-classical light we utilize a Hanbury Brown and Twiss (HBT)-type interferometer, which allows for measuring the second order correlation function (see Figure \ref{fig:2}(c)). We fit a three-level model with excited and meta-stable shelving state to our data. The correlation function is then given by
\begin{align*}
g^{\left(2\right)}\left(\tau\right)=1-Ae^{-\abs{\tau}/t_1}+Be^{-\abs{\tau}/t_2}
\end{align*}
with the anti- and bunching-amplitudes $A$, $B$, and the characteristic lifetimes $t_1$, $t_2$. For the three sample emitters we find $g^{\left(2\right)}\left(0\right)=0.142(37), 0.196(53)$ and $0.234(44)$, respectively. There was no background correction\cite{PhysRevA.64.061802} necessary due to the low detector noise compared to the single-photon brightness. This also means that the observed finite multi-photon probability is not caused by detector dark counts, but rather noise sources excited by the laser. Note that the experimental data was normalized such that for infinite time delay $g^{(2)}(\tau\rightarrow\infty)=1$. As already mentioned, the literature reports ZPLs typically bunch around certain wavelengths. In fact, in our experiments we have seen this to happen around 560$\,$nm, 590$\,$nm and 640$\,$nm, as the histogram in Figure \ref{fig:2}(d) shows. With our fabrication method, however, we were not able to create emitters with ZPLs $>700\,$nm with statistical significance. In addition, sometimes we created an emitter not falling into any of the groups defined above. We believe that these are contaminating fluorescent molecules adsorbed onto the surface of hBN. Their emission is typically much weaker and their spectrum broader compared to the other emitters (see Figure \ref{fig:2}(e)).

\subsection*{Atomic localization of quantum emitters}
\begin{figure*}[t!]
\centering
  \includegraphics[width=\linewidth,keepaspectratio,valign=t]{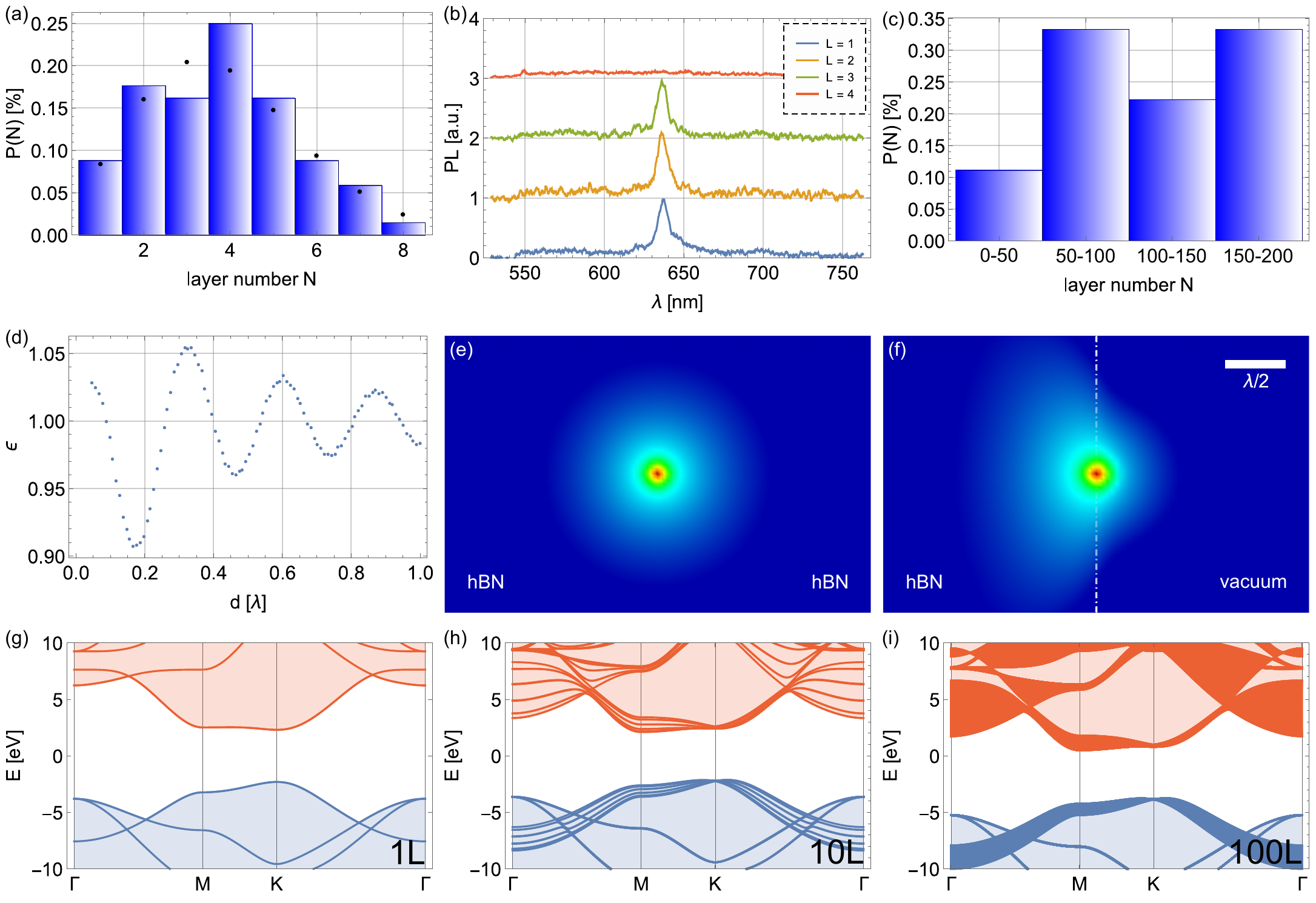}
\caption{Atomic localization of quantum emitters. (a) Probability density of locating the emitters in layer N (i.e.\ it disappeared after N etching steps). The average value is 3.8. The black points are the best fit to any univariate distribution (here a Poisson distribution). The emitters have been created by oxygen plasma treatment. (b) Spectral evolution of one emitter as consecutive layers are removed from the top side. The emission line is relatively stable and suddenly fully disappears after the fourth etching step. (c) Probability density of locating the emitters in layer N (i.e.\ it disappeared after N etching steps). The emitters have been created by electron irradiation. (d) FDTD simulations of the Purcell effect of a dipole emitter close to the hBN-vacuum interface. The emitter lifetime or Purcell factor $\epsilon$ oscillates as the emitter gets moved deeper into the crystal. In the limit of $d\gg\lambda$ there is no enhancement or suppression and the electric field mode profile in this limit is shown in (e). For the limit $d\ll\lambda$ the electric field mode profile is shown in (f). The emission is stronger into the crystal than into the vacuum (as the crystal has a higher dielectric constant). (g-i) DFT calculations of the band structure routed along high-symmetry points for 1L, 10L, and 100L hBN, respectively. Due to layer-layer interactions the bands added by the layers spread, but no deep energy band appears, meaning that the interaction with surface states is likely low.}
\label{fig:3}
\end{figure*}

With 93 quantum emitters fabricated and characterized, we could utilize the atomic etching of hBN, removing one layer at a time. After each cumulative plasma etching step, the flakes were scanned again and we checked if the defect survived (see Supporting Information, Figure S2 for the process flow). It is possible that this etching creates new emitters, but at the layer-by-layer etching parameters, we expect the linear defect formation density to be $\sim 0.02\,\mu$m$^{-1}$ (i.e.\ one defect forms on average per 50$\,\mu$m crystal edge length)\cite{10.1021/acsphotonics.8b00127}. Thus, it is unlikely that an emitter is removed and at the same time a new one forms at the same location. In addition, as the photophysics of the defects vary substantially, it would be even more unlikely that a newly created emitter that formed at the location of a previous emitter has similar photophysical properties (in terms of e.g.\ ZPL, lifetime, and dipole orientation). In fact, we did see occasionally new defects appear at new locations, but they are not counted toward the statistics in this study. The histogram of the number layer after which the defect disappeared is shown in Figure \ref{fig:3}(a). The best fit to any univariate distribution reveals a Poisson distribution with a mean of 3.8. This means that the emitters are very close to the surface.\\
\indent When looking at how the photophysics evolve as the top layers are successively etched, it becomes clear that the emission is stable until the emitter is removed (see Figure \ref{fig:3}(b)). The photoluminescence does not decrease gradually nor change its lineshape. Rather the PL from the defects disappears suddenly entirely, and for all upon removal. This means the quantum emitters are well isolated within one layer with no appreciable inter-layer interaction. In principle, it is possible the wave function of the trapped charge carrier is spread over multiple layers, thus the defect could enter a dark state even if some layers above the layer containing the chemical defect are etched (while the defect itself is not etched yet). There is, however, no further evidence supporting this conjecture. In addition, all the emitters with ZPLs falling not into one of the categories in the histogram in Figure \ref{fig:2}(d) disappeared after the first etching step. This is evidence for the fact that these emitters are indeed surface contaminants. As expected, the Raman shift after each etching step remained constant, indicating that there is not much strain in the crystal, which would relax as the layers are etched.\\
\indent The extracted layer number is believed to be highly accurate. Assuming a failure probability $\leq 0.1\%$ (see above) results in a success probability of 77.2$\%$ that all layer numbers are correct (in total there were 258 etching steps). However, as all samples were etched at the same time, there is a chance that if one process failed, many samples would be affected. A process fail could be that it took a longer time for the plasma to ignite or to stabilize the gases (both ignition and stabilization happens at a higher plasma power, which is subsequently regulated down to the set power), so to exclude this possibility the plasma parameters are recorded \textit{in-situ}.\\
\indent The results so far prove emitters (formed by oxygen plasma treatment) are always very close to the surface. This raises a few questions: (1) Why are the emitters close to the surface? (2) Are emitters always close to the surface, or does this depend on the defect formation method? (3) Is this an explanation for the shorter excited state lifetime of the plasma treated quantum emitters?\\
\indent The dominant ion species in the plasma is O$^{2+}$ (at lower pressure and higher power O$^+$ becomes more dominant). The expected ion energy during the defect formation plasma treatment is $\sim 10\,$eV. Unfortunately, this ion energy is too low for Monte Carlo methods like SRIM\cite{10.1016/j.nimb.2010.02.091}, preventing an accurate calculation of the projected ion range in matter (in this case hBN). However, in our case the plasma treatment is a chemical and not physical process. This means the process is mostly limited to the crystal surface, as the ions have only low kinetic energy and cannot penetrate deep into the crystal. The kinetic energy of the ions is similar to the defect formation energy in hBN, which is on the order of a few eV\cite{PhysRevB.86.245406}. Moreover, the O$_\text{B}$ and O$_\text{N}$ defect have formation energies of 5.19 and 2.20$\,$eV, respectively, so they could easily be produced by the ions\cite{PhysRevB.96.144106}. The oxygen radicals are highly reactive and are thus likely producing defects. It was recently pointed out, however, that it is unclear whether the defects are actually created using the plasma processing or one of the many other methods, or if preexisting, initially dark defects are activated via modification or restructuring of the crystal environment\cite{doi:10.1146/annurev-physchem-042018-052628}. Both options are possible and our data so far does not allow to favor one over the other explanation.\\
\indent While the oxygen plasma only acts onto the crystal surface, defect diffusion is also an important consideration. Without the exact knowledge of the chemical defect structure this is impossible to estimate, but at least a few things are known: First, hBN has strong sp$^2$-hybridized covalent bonds, so the defect diffusion activation energy (that is the energy required to move along the reaction path) is rather large. It is expected that diffusion is predominantly in-plane and not inter-layer due to the direct in-plane bonds, so diffusion deep into the crystal is not likely. For hBN, due to the heteronuclear structure, defect diffusion is partially suppressed, as homonuclear B-B and N-N are energetically unfavorable (these homonuclear bonds are temporarily formed as the defects moves along the reaction path)\cite{PhysRevB.75.094104}. This reduces e.g.\ vacancy migration compared to graphene. The diffusion activation energy calculated with DFT range from 2.6 to 6.0$\,$eV at 0$\,$K for vacancies and divacancies, with the structures often relaxing to their initial configuration\cite{PhysRevB.75.094104}. This already shows the smaller defect diffusion. Furthermore, at the rapid annealing temperature of 850$^\circ$C (in this experiment), only the boron vacancy has a diffusion coefficient larger than 1$\,$\AA$^2$s$^{-1}$\cite{PhysRevB.75.094104}. Future calculations have to show how the diffusion of other point-like complexes scales. It is worth noting, that the result of the defect diffusion activation energy from DFT calculations shows a small dependency on the specifically used pseudopotential\cite{doi:10.1002/anie.201100733}.\\
\indent To address the second question, we repeat the experiment with emitters fabricated with electron irradiation\cite{10.1021/acsami.6b09875,10.1021/acsami.8b07506}. The electron accelerating voltage was 10$\,$kV with an electron fluence of $\sim 10^{18}\,$cm$^{-2}$. Given the thickness of the hBN flakes being $\ll 1\,\mu$m, the kinetic energy of the electrons is sufficient to fully transmit through the hBN crystals (see Supporting Information, Figure S3(a)). The energy loss of the electrons is dominated by collisions with the boron and nitrogen nuclei, as the radiative stopping power is much smaller at 10$\,$keV kinetic electron energy (see Supporting Information, Figure S3(b)). Therefore, bremsstrahlung does not play any role. With the projected range of the electrons being 1.4$\,\mu$m at $10\,$keV, it is expected that emitters created or activated by electron irradiation are not exclusively near the crystal surface. Monte Carlo simulations of electron trajectories through the hBN crystal (see Supporting Information, Figure S3(c,d)) also confirm this. Repeating the atomic etching on these new emitters confirms this, as none of the emitters was found within the first ten layers, and the emitters being randomly positioned within the crystal. Etching at much larger steps ($\sim 10$s of layers at a time, even though we note this was not calibrated sufficiently) shows that defects created by electron irradiation are formed throughout the crystal (see Figure $\ref{fig:3}$(c)). More precisely, the emitters form not exclusively at the crystal edges or dislocations anymore, in agreement with previous experiments\cite{10.1021/acsami.8b07506}. Interestingly, the excited state lifetime of these emitters is typically longer compared to the plasma etched ones, with lifetimes ranging from $2-3\,$ns (see Supporting Information, Figure S4).

\subsection*{Theoretical modeling}
Finally, we address the third question. Within the crystal, the photon density of states is decreased compared to vacuum. This is a Purcell-like effect, where the radiative lifetime is modified as the dielectric environment changes. The Purcell factor $\epsilon$ as a function of emitter distance to the surface $d$ is calculated using finite-difference time-domain (FDTD) simulations (see Methods) and shown is in Figure \ref{fig:3}(d). The Purcell factor (and thus the excited state lifetime of an ideal dipole) oscillates and reaches 1 in the limit $d\gg\lambda$. In this limit there is no enhancement or suppression. It becomes clear that this effect only makes up a few percent in lifetime changes, so this alone cannot explain the shorter lifetime. It is still noteworthy, that there is enhancement very close to the surface, while deeper ($45-145\,$nm) there is suppression. The electric field mode profiles in both limits show the emitter deep within hBN emits like an ideal dipole, while the emitter at the surface emits stronger into the crystal than into vacuum (see Figure \ref{fig:3}(e,f)). This means the actual emitter brightness is even larger than experiments so far suggest. For emitters in cavities\cite{arXiv:1902.03019}, this does not matter, as both directions are captured by the cavity. As the different lifetime is not solely due to a Purcell-like effect, we use density functional theory calculations to investigate if surface states could be the cause for the shorter lifetime. We calculate the electronic band structure of hBN for one (1L), ten (10L), and 100 layers (100L) of hBN (see Figure \ref{fig:3}(g-i)). The calculations show, that as more layers are added also more energy bands are added. Due to layer-layer interactions these bands spread, but there are no genuine isolated surface bands introduced into the band gap. This implies that, unless the defect levels are very close to one of the band edges, surface states do not influence the lifetime of the defect. Therefore, we conclude that the shorter defect lifetime in our experiments is likely due to interaction with surface defects introducing additional decay pathways, or with surface phonons making existing decay pathways faster.

\section*{Conclusions}
In this work, we have developed deterministic atomically layer-by-layer etching of hBN with an oxygen plasma. This was utilized to destructively localize quantum emitters hosted by hBN. We found that emitters fabricated by a different plasma process are always very close to the surface, within a few layers, while emitters fabricated by intense electron irradiation are located throughout the entire crystal thickness. For both creation methods, emitters are more likely to form at flake edges and grain boundaries. It is notable that they also form away from these domains, in what appears to be undistorted crystal. Creation near the surface is a likely explanation for the shorter excited state lifetime hBN quantum emitters exhibit when fabricated by plasma etching. The emitter lifetime is influenced by additional decay pathways introduced by surface defects, or interactions with surface phonons making existing decay pathways faster. In contrast, emitters deep within the crystal have lifetimes $\sim 3-6$ times longer, as they are well isolated from the environment and surface effects.\\
\indent Considering now the implications that our observations have for the identity of the quantum emitters. Our etching study is consistent with the confinement of the emitting defect to a single layer, as per past observation of the emitters in monolayer samples\cite{nnano.2015.242}. The creation of deep defects away from a boundary by electron irradiation is an important observation. It implies that the defect can be a product of radiation damage and so is further evidence that it involves a vacancy or interstitial. Specifically either a nitrogen vacancy V$_\text{N}$, a boron-vacancy V$_\text{B}$, an intralayer interstitial or an interlayer interstitial. To identify which, we need to interpret the effects of annealing.\\
\indent At our annealing temperature of 850$^\circ$C, it is known that the V$_\text{B}$ is mobile, whilst V$_\text{N}$ is not. It is reasonable to expect that the interlayer interstitials are also mobile due to the low interlayer bond energies of the material. Upon annealing, we observe improved photostability and linewidth, but no significant change in the number of emitters\cite{10.1021/acsphotonics.8b00127}. We attribute the improved optical properties to the removal of interstitials and single V$_\text{B}$, which we expect to lead to an improved charge stability and reduced electrical noise since these defects likely act as donors or acceptors. If the density of the V$_\text{B}$ created by the radiation is low, then our annealing observation would imply that the defect does not involve V$_\text{B}$. This is because if the defects were single V$_\text{B}$, then the number of emitters would decrease with annealing, and if it were a complex involving one or more V$_\text{B}$, then the number of emitters would increase until saturation of the other constituents of the complex (i.e.\ V$_\text{N}$ or impurities). However, we are not necessarily drawing this conclusion here, since our intense electron irradiation may have rather created a very high density of V$_\text{B}$, which even without annealing, could have saturated the creation of emitters (i.e.\ by creating V$_\text{B}$ in close proximity to V$_\text{N}$ or an impurity). In this case, the defect may well involve V$_\text{B}$. Unfortunately, we cannot determine which V$_\text{B}$ density limit our radiation produced because there is insufficient information about the V$_\text{B}$ creation cross-section for electron radiation. Future work should focus on establishing the V$_\text{B}$ density created before annealing and relating this to the creation / destruction / no change of emitters during annealing to establish whether or not V$_\text{B}$ is involved in the defect.\\
\indent The creation of the defects by the oxygen plasma may imply that the defect involves oxygen impurities through their incorporation at the surface. Generalizing this hypothesis to the creation of deep emitters by electron irradiation, this would imply that oxygen is also a deep impurity in our samples. This appears reasonable given that O may also form similar sp$^2$ bonds as B and N if it can donate an electron to a nearby acceptor. Future work should seek to combine variation of oxygen impurity and radiation damage to ascertain whether the defect is indeed an O-V complex.\\
\indent The results might also allow for a direct identification of the defect, as the knowledge of optically active defects very close to the surface might allow for imaging with high-resolution tunneling electron microscopes. A full understanding of the defect nature is required for tuning and engineering specific properties that will ultimately lead to a wider applicability in various scenarios.

\section*{Methods}
\subsection*{Plasma etching}
The oxygen plasma was generated from a microwave field (PVA TePla). Prior to any experiments, the plasma chamber was cleaned for 5$\,$min at 500$\,$W to remove any contaminants. We found the optimal single layer etching conditions empirically at a plasma power of 102$\,$W for 63$\,$s at a pressure of 0.332$\,$mbar and a gas flow rate of 300$\,$cm$^3$/min (deviating from this by 10$\%$ decreases the success probability). All experiments were carried our at room temperature. The plasma time includes about $2-3\,$s during which the plasma ignites and the gases are stabilized. The plasma field is highly anisotropic and varies across the plasma chamber. Thus, for repeatable results it is crucial to place the substrates always at the same position in the chamber. Unless stated otherwise, this position is at the plasma field minimum. It should be mentioned that the optimal parameters reported here depend on the specific gas pump, plasma generator and geometry of the chamber, which requires to optimize these parameters on every other system individually.
\subsection*{Fabrication and optical characterization}
Thin flakes of hBN were mechanically exfoliated from bulk crystal (used as received from HQGraphene) to a viscoelastic stamp (Gel-Pak WF-40-X4) using the  tape method. Crystals with thicknesses down to $\sim 5\,$nm can be identified by optical contrast with a standard optical microscope and are subsequently transferred by dry contact to a Si substrate with a 262$\,$nm thermally grown oxide layer. For the quantum emitters, we used crystals with thicknesses ranging from $\sim 5-100\,$nm. The emitters were created during an oxygen plasma etching step at 200$\,$W in the plasma field maximum and subsequently rapidly thermally annealed at 850$\,^\circ$C in an Ar atmosphere. The electron irradiated emitters have been fabricated using a scanning electron microscope in an FEI Helios 600 NanoLab, where the electrons were accelerated using a high voltage of 10$\,$kV. The samples were irradiated with a fluence of $f=10^{18}\,$cm$^{-2}$, which was calculated with $f=\frac{I\cdot t}{e\cdot A}$, where $I$ is the electron current, $t$ is the frame time, $e$ is the electron charge, and $A$ is the frame area. The irradiation took place at room temperature at a pressure $<2.2\,$mPa. For emitter localization, a custom-built $\mu$PL setup was used which utilized an ultrashort-pulsed 522$\,$nm laser with a pulse length of 300$\,$fs at a repetition rate of 20.8$\,$MHz. The laser was focused to the diffraction limit with a Olympus $100\times$/0.9 dry objective and the samples were scanned using Newport translation stages with a spatial resolution up to 0.2$\,\mu$m. The emission was collected in-reflection through the same objective and frequency-filtered using Semrock RazorEdge ultrasteep long-pass edge filters. The light is coupled via a grating to either a CCD or a single-photon avalanche diode (SPAD) from Micro Photon Devices allowing to extract the spectrum or the temporally and spectrally resolved photoluminescence. The correlation between excitation pulse and arrival time of the fluorescence photon is given by a PicoHarp 300. For measuring the second-order correlation function we utilize another diode laser at 512$\,$nm and two SPADs.
\subsection*{Finite-difference time-domain simulations}
The finite difference time-domain simulations were performed using Lumerical FDTD Solutions, a commercial grade simulator based on the FDTD method\cite{Lumerical}. To calculate the Purcell enhancement and emitter dynamics, an in-plane dipole emitter at 560$\,$nm was defined in the center within a slab of hBN, with a dielectric constant of 3.42 at 532$\,$nm (this was obtained from experiments). The slab was thinned down from one direction (which is equivalent to moving the emitter to the surface) and the Purcell enhancement as well as the electric field mode profile was recorded for each crystal thickness. A dynamic mesh was chosen to capture all potential emitter dynamics. The simulations assume perfectly matched layer boundary conditions, which are reflectionless or absorbing boundaries, to account for the finite memory size.
\subsection*{Density functional theory calculations}
The DFT calculations have been performed with QuantumATK with the Virtual NanoLab front end\cite{PhysRevB.96.195309,ATK}. QuantumATK utilizes numerical linear combination of atomic orbitals basis sets and the density matrix for closed or periodic systems is calculated by diagonalization of the Kohn-Sham Hamiltonian. Monolayer hBN crystals have been defined using a supercell containing two atoms and the geometry has been optimized using a $21\times 21\times 1$ Monkhorst-Pack reciprocal space grid. The optimization converged when all forces were below $0.001\,$eV$\,$\AA$^{-1}$. The electron exchange-correlation was described with the Perdew-Burke-Ernzerhof (PBE) functional in the generalized gradient approximation\cite{PhysRevLett.77.3865}. For all atoms a double zeta polarized basis set was chosen and band structure was routed along high symmetry points. The ten- and 100-layer hBN crystals have been constructed in a similar way, with the lattice constant $c$ also geometrically optimized and the k-sampling in this direction chosen such that it does not influence the simulation results.
\begin{acknowledgement}
This work was funded by the Australian Research Council (CE170100012, FL150100019, DE140100805, DP180103238, DE170100169). We thank the ACT Node of the Australian National Fabrication Facility for access to their nano- and microfabrication facilities. We also thank H. Tan for access to the TRPL system and C. Corr and F. Karouta for useful discussions about plasma processing.
\end{acknowledgement}

\providecommand{\latin}[1]{#1}
\makeatletter
\providecommand{\doi}
  {\begingroup\let\do\@makeother\dospecials
  \catcode`\{=1 \catcode`\}=2 \doi@aux}
\providecommand{\doi@aux}[1]{\endgroup\texttt{#1}}
\makeatother
\providecommand*\mcitethebibliography{\thebibliography}
\csname @ifundefined\endcsname{endmcitethebibliography}
  {\let\endmcitethebibliography\endthebibliography}{}

\clearpage

\clearpage
%

\end{document}